\documentclass[twocolumn,showpacs,amsmath,nofootinbib,pra,aps,amssymb]{revtex4-1}

\usepackage[T1]{fontenc}
\usepackage[utf8]{inputenc}
\usepackage{times}
\usepackage{color} 
\usepackage{array}
\usepackage{amssymb,amsmath}
\usepackage{amsbsy}
\usepackage[pdftex]{graphicx}
\usepackage{bm}
\usepackage{float}
 
\usepackage[unicode,breaklinks]{hyperref}
\hypersetup{
    unicode=true,
    plainpages=false, 
    colorlinks=true,
    linkcolor=blue,
    citecolor=blue,
    filecolor=black,
    urlcolor=blue
}
\usepackage{url}
\usepackage{verbatim}

\newcommand{\be}{\mathbf{e}}
\newcommand{\bk}{\mathbf{k}}
\newcommand{\br}{\mathbf{r}}
\newcommand{\bx}{\mathbf{x}}
\newcommand{\bB}{\mathbf{B}}
\newcommand{\kh}{\hat{k}}
\newcommand{\xh}{\hat{x}}
\newcommand{\yh}{\hat{y}}
\newcommand{\zh}{\hat{z}}
\newcommand{\UD}{U_{dd}}
\newcommand{\LGP}{\mathcal{L}_\mathrm{GP}}
\newcommand{\UDt}{\tilde{U}_{dd}}

\begin{document}

\title{Rotational tuning of the dipole-dipole interaction in a Bose gas of magnetic atoms}
\author{D.~Baillie}  
\affiliation{Dodd-Walls Centre for Photonic and Quantum Technologies, New Zealand}
\affiliation{Department of Physics, University of Otago, Dunedin 9016, New Zealand}
\author{P.~B.~Blakie}  
\affiliation{Dodd-Walls Centre for Photonic and Quantum Technologies, New Zealand}
\affiliation{Department of Physics, University of Otago, Dunedin 9016, New Zealand}

\begin{abstract}  
We investigate the dynamics of a Bose-Einstein condensate of magnetic atoms in which the dipoles are rotated by an external magnetic field. The time-averaged dipole-dipole interaction between the atoms is effectively tuned by this rotation, however recent experimental and theoretical developments show that dynamic instabilities emerge that may cause heating. We present simulations of a realistic tuning sequence in this system, and characterize the system behavior and the emergence of instabilities. Our results indicate that the instabilities develop more slowly as the rotation frequency increases, and indicate that experiments with tuned dipole-dipole interactions should be feasible.
\end{abstract} 

\maketitle

\section{Introduction}
The ability to manipulate the short-ranged interactions in ultra-cold quantum gases using Feshbach resonances \cite{Chin2010a} has enabled a wide range of quantum many-body physics to be investigated \cite{Bloch2008a}. With the production of quantum degenerate gases of highly magnetic atoms (e.g.~Cr \cite{Griesmaier2005a}, Dy \cite{Mingwu2011a} and Er \cite{Aikawa2012a}) the field of quantum gases now has access to systems with long ranged dipole-dipole interactions (DDIs).
 It is desirable to tune the magnitude and sign of the DDI independently of the short-ranged interaction. Being able to do so would allow experiments to explore new phenomena, for example, stabilizing two-dimensional bright solitons \cite{Pedri2005a}, realizing rotonic Kelvin-wave excitations on vortex lines \cite{Klawunn2008a}, or allowing the interactions between vortices to be manipulated \cite{Mulkerin2013a} (also see \cite{Lahaye_RepProgPhys_2009}). In 2002, Giovanazzi {\sl et al.}~\cite{Giovanazzi2002b} proposed a scheme for tuning the DDIs by rotating the dipoles fast enough that the interaction can be time-averaged.   This scheme was recently implemented in an experiment with a Bose-Einstein condensate (BEC) of $^{162}$Dy atoms by Tang \textit{et al.}~\cite{Tang2018a}. In these experiments, the anisotropic expansion dynamics of the condensate (which depends on the magnitude and sign of the DDI) was used to reveal the effects of the tuned interactions.

In practice, DDI tuning is implemented by rotating the magnetic field (used to polarize the atoms) about an axis at frequency of $\Omega$. If  $\Omega$ is smaller than the Larmor frequency $\omega_{\mathrm{L}}$ (typically $\omega_{\mathrm{L}}\gtrsim10^6\,$Hz), then the dipoles will follow the magnetic field. It has been assumed that if this rotation is fast compared to the trap frequencies $\omega_{\mathrm{tr}}$ (typically $\omega_{\mathrm{tr}}\sim10^2\,$Hz), and hence the typical timescale of atomic motion, then the rotating dipoles can be time-averaged to describe their effect on the condensate. The angle $\varphi$ of the magnetic field with respect to its axis of rotation, determines the strength of the time-averaged DDI: the time-averaged DDI takes the form of a static DDI for dipoles polarized along the axis of rotation, but with the bare DDI coupling constant $g_{dd}$ scaled by a factor of $\frac12(3\cos^2\varphi-1)$. Thus, by varying  $\varphi$, the time-averaged DDI  can be reduced in strength, made zero,  or even negative \cite{Giovanazzi2002b,Lahaye_RepProgPhys_2009}.

 The experiments of Tang \textit{et al.}~\cite{Tang2018a} observed that when tuning the DDI,  the lifetime of their condensate was reduced by more than an order of magnitude. They speculated that the configuration of coils used in the experiment, which  generated residual magnetic field gradients, could have contributed to this lifetime reduction.
  However, with a condensate lifetime of $\sim 160\,$ms, they concluded it would be sufficiently long for many types of experiments with tuned DDIs. 

 A recent theoretical treatment of the meanfield dynamics of a condensate with a rotating DDI was presented by Prasad \textit{et al.}~\cite{Prasad2018a}. Their analysis, performed in the Thomas Fermi limit (where the zero-point kinetic energy terms are neglected), predicted that with rotating dipoles the condensate was dynamically unstable in a broad parameter regime.  
 Significantly, they concluded that this instability prevents the formation of a stable long-lived rotationally-tuned BEC. Their work was performed in the co-rotating frame of the magnetic dipoles and did not directly model the kind of dynamical scenario used in experiments to tune the DDIs. They present a dynamical simulation demonstrating the instability, but for a rather slow rotation frequency case ($\Omega=3\omega_{\mathrm{tr}}$), notably with $\Omega<\mu/\hbar$, where $\mu$ is the condensate chemical potential that usefully quantifies the interaction energy scale.
So, important questions remain about the instability time-scale for experimentally relevant tuning processes, and how the instability time depends on the rotation frequency relative to characteristic frequencies of the condensate.

 Here we address these questions by simulating the instability dynamics of a dipolar BEC for a realistic tuning scenario. We consider the case of magnetic dipoles taken into a tuned DDI configuration from an initially stationary (untuned) state using an appropriate ramping procedure. Our simulations are based on the truncated Wigner formalism using a nonlocal Gross-Pitaevskii equation (GPE) with noise added to simulate quantum fluctuation effects. We use measures of width and the DDI energy to quantify the tuning of the dipole interactions in our simulations.
 In general we find that this ramp excites collective dynamics of the condensate, but this can be minimized for ramp times exceeding the characteristic trap period.  We identify the kinetic energy as a good observable to reveal the onset of dynamic instabilities arising from the DDI tuning. We find that increasing  rotation frequency of the dipoles delays the onset of instability.  
 Our results indicate that DDI tuning with minimal heating over long time scales ($\gtrsim100\,$ms) should be feasible in experiments with sufficiently fast rotation frequencies.
 
\section{Theory}
\subsection{Formalism}
We consider a uniform external magnetic field aligned at polar angle $\varphi(t)$ to the $z$ axis and rotating at rate $\Omega$, i.e. 
$\bB(t) = B\be(t)$ where 
\begin{align}
    \be(t) = (\xh\cos\Omega t+\yh\sin\Omega t)\sin\varphi(t) + \zh\cos\varphi(t).\label{eq:e(t)}
\end{align}
We assume that $\Omega \ll \omega_{\mathrm{L}}=\mu_m B/\hbar$, where $\mu_m$ is the magnetic moment of the atoms, so that magnetic dipoles follow the external field, i.e. the dipoles are aligned along $\be(t)$. The time-dependent GPE for the wavefunction $\psi$ is $i\hbar\dot{\psi} = \LGP\psi$, with 
\begin{align}
    \LGP = H_{sp} + g_s|\psi|^2 + \Phi_D(\bx,t), \label{LGP}
\end{align}
where 
\begin{align}
    H_{sp} = -\tfrac{\hbar^2}{2m} \nabla^2  + \tfrac{m}{2} (\omega_x^2x^2+\omega_y^2y^2+\omega_z^2z^2),
\end{align}
is the single particle Hamiltonian.  
The atoms interact by a contact interaction with coupling constant $g_s=4\pi\hbar^2a_s/m$, where $a_s$ is the $s$-wave scattering length (which we take to be positive). They also interact by a DDI described by the potential \cite{Lahaye_RepProgPhys_2009}
\begin{align}
    {\UD}(\br,t) &= \frac{3{g}_{dd}}{4\pi }\frac{1-3[\be(t)\cdot\hat{r}]^2}{r^3},\label{e:Ur}
\end{align}
where  $g_{dd}=\mu_0 \mu_m^2/3\ge0$ is the  DDI coupling constant. In Eq.~(\ref{LGP}) the DDI appears via the effective
 dipolar interaction potential $\Phi_D(\bx,t) = \int d\bx'\,\UD(\bx-\bx',t)|\psi(\bx')|^2$ (we omit the $t$ dependence of $\psi$).
The energy of the system is
\begin{align}
    E[\psi] = \int d\bx\, \psi^*\left[H_{sp} + \frac12g_s|\psi|^2+ \frac12 \Phi_D(\bx,t) \right]\psi. \label{Efunc}
\end{align} 
The strength of the DDIs is also conveniently characterized in terms of the dipole length $a_{dd}\equiv m\mu_0\mu_m^2/12\pi\hbar^2$, such that $g_{dd}=4\pi a_{dd}\hbar^2/m$. The dimensionless number
$\epsilon_{dd}\equiv{g_{dd}}/{g_s}={a_{dd}}/{a_s}$ describes the ratio of the (untuned)  DDIs to the contact interactions. 

\subsection{Time-averaged DDI}
For $\varphi(t)=\varphi$ fixed and $\Omega$ sufficiently high we can time-average the dipole motion to obtain the averaged DDI
\begin{align}
   \bar{U}_{dd}(\br) &= \frac{3\bar{g}_{dd}(\varphi)}{4\pi }\frac{1-3\cos^2\theta}{r^3},\label{e:Urav}
\end{align}
where $\theta$ is the angle between $\br$ and the $z$ axis and the effective time-averaged coupling constant is \cite{Giovanazzi2002b}
\begin{align}
    \bar{g}_{dd}(\varphi) &= g_{dd} \left(\frac{3\cos^2\varphi-1}2\right).
\end{align}
By choice of the angle $\varphi$, we can vary the time-averaged coupling constant from $\bar{g}_{dd}(0)=g_{dd}$  to $\bar{g}_{dd}(\frac{\pi}{2})=-\frac{1}{2}g_{dd}$.

Stationary states $\bar{\psi}$ of the time-averaged Hamiltonian satisfy 
\begin{align}
\bar{\mu}\bar{\psi}=\bar{\mathcal{L}}_{\mathrm{GP}}\bar{\psi},
\end{align}
with $\bar{\mathcal{L}}_{\mathrm{GP}} = H_{sp} + g_s|\bar{\psi}|^2 + \bar{\Phi}_D(\bx,t)$, and $\bar{\Phi}_D=\int d\mathbf{x}'\,\bar{U}_{dd}(\mathbf{x}-\mathbf{x}')|\bar{\psi}(\mathbf{x}')|^2$. 

\section{Results\label{s:results}}

\begin{figure}[htbp] 
   \centering
   \includegraphics[width=3.4in]{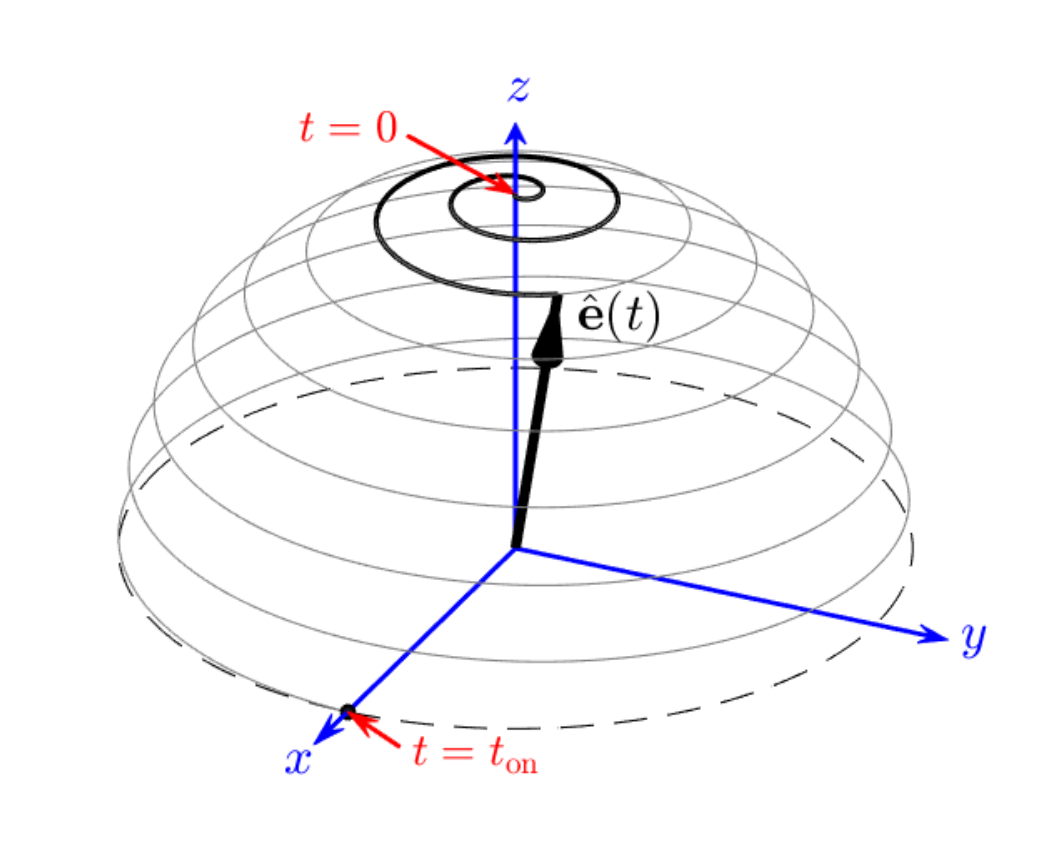}
   \caption{Schematic of the dipole tuning ramps we use in our simulations. The dipole direction $\mathbf{e}$ starts along $z$ and then spirals down, using a linear ramp of the polar angle $\varphi$, over a time $t_{\mathrm{on}}$ into the $xy$-plane according to Eq.~(\ref{eq:e(t)}). For $t>t_{\mathrm{on}}$ the dipole executes uniform circular motion in the $xy$-plane with an angular velocity $\Omega$.
   }
   \label{fig:schematic}
\end{figure}

We consider the dynamics of a dipolar BEC taken from an initial condition of a static dipole oriented along $z$, into a configuration with the dipoles rotating in the $xy$-plane $\varphi=\tfrac{\pi}{2}$, for which $\bar{g}_{dd}=-\tfrac{1}{2}g_{dd}$. We do this by maintaining a constant angular frequency of rotation $\Omega$ about the $z$ axis, and  tilt the dipole by linearly increasing $\varphi$ from $0$ to $\pi/2$ over the time period $t_{\mathrm{on}}$ (see Fig.~\ref{fig:schematic}). We work in the regime that $\Omega\gg 1/t_{\mathrm{on}}$, so that it is reasonable for analysis to time average over the rotation to obtain an effective interaction $\bar{g}_{dd}$ that varies with $\varphi$ on the slower timescale of $t_{\mathrm{on}}$.

For our simulations the initial BEC is taken as the ground state  $\psi_0$ of the static configuration, i.e.~$\mu \psi_0 = \mathcal{L}_{\mathrm{GP}}\psi_0$ with $\varphi=0$, and $N_c=\int d\mathbf{x}\,|\psi_0|^2$ condensate atoms. We add noise to the ground state to account for the effects of quantum fluctuations, which can seed instabilities in the system dynamics. This noise is added as half an atom per mode (on average) in the single particle harmonic oscillator basis up to a single particle energy cutoff of $\epsilon_{\mathrm{cut}}$. Formally this method of adding noise corresponds to the truncated Wigner prescription (see \cite{cfieldRev2008}).
This  state is then evolved according to the GPE $i\hbar\dot{\psi}=\mathcal{L}_{\mathrm{GP}}\psi$ using the time-dependent  dipole polarization outlined above.  More details about the initial state preparation and evolution are given in Appendix \ref{AppNum}.

\begin{figure}[htbp] 
   \centering
   \includegraphics[width=3.4in]{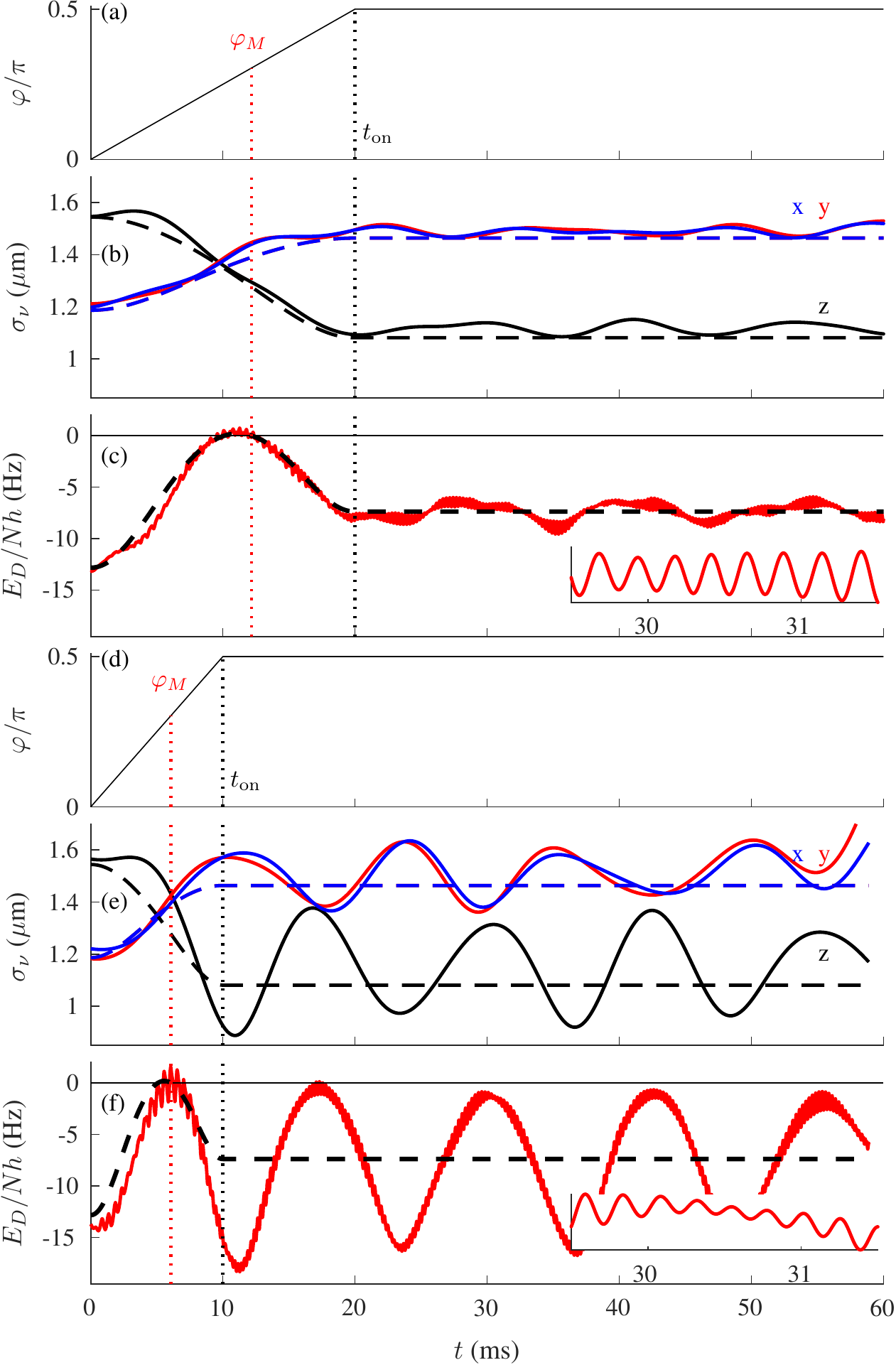}
   \caption{System dynamics arising from a dipole tuning ramp with $\Omega/2\pi=2\,$kHz. Results are given for  $t_{\mathrm{on}}=20\,$ms (a,b,c) and  $t_{\mathrm{on}}=10\,$ms  (d,e,f) ramps, with the  polar angle shown in (a) and (d). The evolution of the system widths (b,e) and dipole energy (c,f).  In (b) and (e) the instantaneous widths (solid lines: red $\sigma_x$, blue $\sigma_y$ and black $\sigma_z$) are calculated from the time-dependent GP simulation $\psi$. We also show the ground state $\bar{\psi}$ widths  [dashed lines: red $\bar\sigma_x$, blue $\bar\sigma_y$ and black $\bar\sigma_z$, with $\bar\sigma_\nu=(\int d\mathbf{x}\,\nu^2|\bar\psi|^2/N)^{1/2}$]  for the time-averaged interaction at the current value of $\varphi$. 
  In (c) and (f) we show the instantaneous dipole energy $E_D$ (solid red line) and the value of the time-averaged ground state $\bar{E}_D$ (dashed black line). 
   The results are for a condensate of $N_c=2.5\times10^3$ $^{164}$Dy atoms, with $a_{dd}=130.8\,a_0$ and $a_s=174.4\,a_0$ (i.e.~$\epsilon_{dd}=0.75$), confined in a harmonic trap with $(\omega_x,\omega_y,\omega_z)/2\pi=(50,50,55)\,$s$^{-1}$. The initial state chemical potential is $\mu=h\times 264\,$ Hz. Noise is added with $\epsilon_{\mathrm{cut}}=2\mu$.
   }
   \label{fig:widthsED}
\end{figure}

We can monitor the effect of the dipole dynamics on the density distribution through the evolution of its widths, which we characterize  by the second moment of the position coordinates, i.e.~$\sigma_\nu=(\int d\mathbf{x}\,\nu^2|\psi|^2/N)^{1/2}$, for $\nu=\{x,y,z\}$, where $N$ is the norm of $\psi$. 
The results in Fig.~\ref{fig:widthsED}(b) show the width dynamics for a  ramp with $t_{\mathrm{on}}=20\,$ms and $\Omega/2\pi=2\,$kHz.  The widths change significantly during the dipole ramp in a manner that is consistent with the system experiencing an effective dipole interaction of $\bar{g}_{dd}(\varphi)$. To see this, we compare the widths from the dynamic simulation to the those of the ground state $\bar{\psi}$ of the time-averaged interaction (dashed lines), and find good quantitative agreement.  Because the ramp time is comparable to the trap period we find that  collective modes are excited during the ramp, as revealed by the oscillating widths that persist after the ramp has concluded.

The width behavior can be understood by considering the DDI energy  
$E_D=\frac{1}{2}\int d\bx\,\Phi_D|\psi|^2$. Because the DDI is anisotropic, the density distribution, and  importantly its aspect ratio, can change to reduce the DDI energy --  a phenomenon generally referred to as magnetostriction \cite{Tang2018a}. The dynamics of $E_D$  are shown in Fig.~\ref{fig:widthsED}(c). We note that since $E_D$ is evaluated using the instantaneous DDI potential it has small and rapid oscillation at a frequency of $2\Omega$ (see inset) about a mean value that varies on the time-scale of the ramp.
For comparison we also show the dipole energy for the ground state of the time-averaged interaction [i.e.~$\bar{E}_D=\frac{1}{2}\int d\bx\,\bar{\Phi}_D|\bar\psi|^2$,  dashed line in  Fig.~\ref{fig:widthsED}(c)] and find that it is similar to the dynamical result.  This confirms that rotating the dipoles can effectively tune the DDIs in the system.

 At the start of the ramp $\varphi=0$, so the bare and time-averaged interactions are identical, $\bar{g}_{dd}=g_{dd}$, and the density distribution is prolate ($\sigma_z>\sigma_{x,y}$), with a negative value of $E_D$.  At $t\!\approx
 \!12.2\,$ms the polar angle is at the magic value $\varphi_M=\cos^{-1}\!\frac{1}{\sqrt{3}}\approx54.7^\circ$, where the time-averaged DDI is zero and the density distribution is close to isotropic, with $E_D\approx0$.\footnote{In our results the time-averaged ground state becomes isotropic, and $E_D$ is maximized,  prior to the magic angle because the harmonic trap is  oblate.} At the conclusion of the ramp $\bar{g}_{dd}=-\frac{1}{2}g_{dd}$ and the  energy is minimized by the density distribution being oblate  ($\sigma_z<\sigma_{x,y}$).
 
 In Fig.~\ref{fig:widthsED}(d)-(f) we consider a faster ramp of $t_{\mathrm{on}}=10\,$ms that is shorter than the trap period. In this case, the collective modes are more strongly excited. The out of phase oscillation of $\sigma_z$ and $\sigma_{x,y}$ [see Fig.~\ref{fig:widthsED}(e)] reveals that the dominant mode excited has a quadrupolar character. This also manifests as an oscillation in the dipole energy. For this ramp time  we consider the condensate to be strongly excited as the amplitude of the energy oscillation in $E_D$ is comparable to the expected mean value $|\bar{E}_D|$ for the time-averaged ground state. 
  From hereon we focus on the slower  $t_{\mathrm{on}}=20\,$ms ramp to ensure that the state prepared at the conclusion of the ramp is close to the time averaged ground state.

\begin{figure}[htbp] 
   \centering
   \includegraphics[width=3.4in]{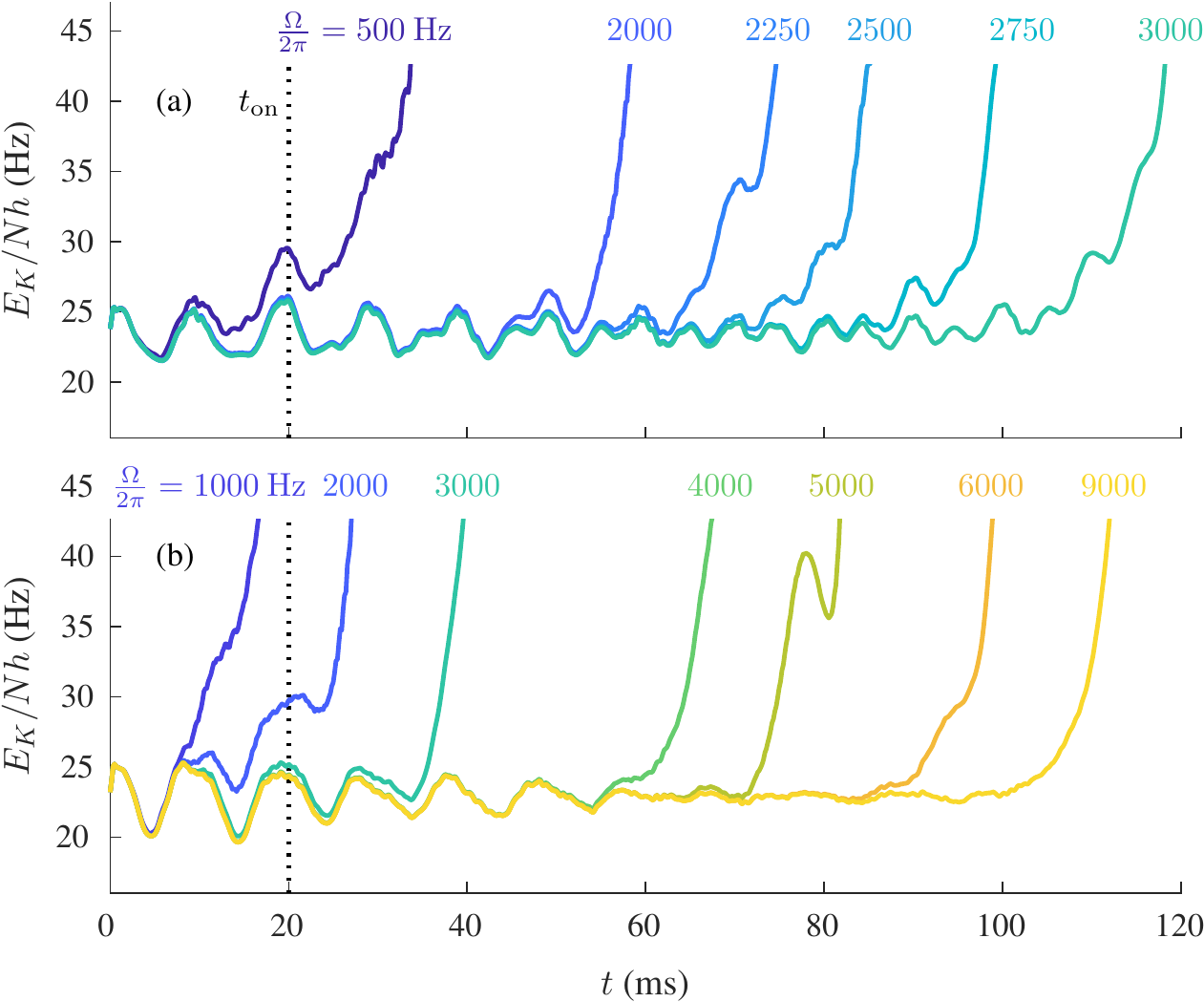} 
   \caption{Dynamics of kinetic energy  for  $t_{\mathrm{on}}=20\,$ms for a condensate with (a) $N_c=2.5\times10^3$ atoms giving $\mu=h\times 264\,$Hz and (b) $N_c=5\times10^3$ atoms giving $\mu=h\times 342\,$Hz. Labels differentiate trajectories with different rotation frequencies $\Omega$. All trajectories  for each subplot  use the same initial state.
   Other parameters are as in Fig.~\ref{fig:widthsED}.
   }
   \label{fig:dynamics1}
\end{figure}

We now consider the emergence of dynamical instabilities after the dipole ramp.  We find that these instabilities are revealed by monitoring the kinetic energy 
\begin{align}
E_K=-\tfrac{\hbar^2}{2m}\int d\bx\,\psi^*\nabla^2\psi.
\end{align}
 In Fig.~\ref{fig:dynamics1}(a) we show the evolution of $E_K$  for various rotation frequencies $\Omega$. Initially  $E_K$  displays oscillatory dynamics at a frequency comparable to the trap frequency, arising from the collective modes excited. However, at later times $E_K$ suddenly starts rapid growth, corresponding to the instability identified in \cite{Prasad2018a} and marking where the condensate begins to heat.  
For the $\Omega/2\pi=2\,$kHz case, corresponding to the same simulation shown in Fig.~\ref{fig:widthsED}(a)-(c),   $E_K$ is seen to start growing at $t\approx50\,$ms and appears to strongly diverge by $60\,$ms, even though little evidence of the instability is apparent at these times in either the widths or dipole energy [see Fig.~\ref{fig:widthsED}(b)-(c)].  For simulations with faster rotation frequencies the initial phase of oscillatory $E_K$ dynamics is almost identical (independent of $\Omega$), however the transition to unstable behavior is seen to be delayed to later times. For $\Omega/2\pi=3\,$kHz, the instability does not emerge until $t\approx110\,$ms.  

We repeat the above analysis for a condensate with twice as many atoms ($N_c=5\times10^3$) in Fig.~\ref{fig:dynamics1}(b). Here we see that for the same value of $\Omega$ the dynamic instabilities tend to manifest at earlier times than for the lower atom number case of Fig.~\ref{fig:dynamics1}(a). Nevertheless, for sufficiently high rotation frequencies (here $\Omega/2\pi>8\,$kHz), the instability can be pushed out to $t\gtrsim100\,$ms.

 \begin{figure}[htbp] 
   \centering
   \includegraphics[width=3.4in]{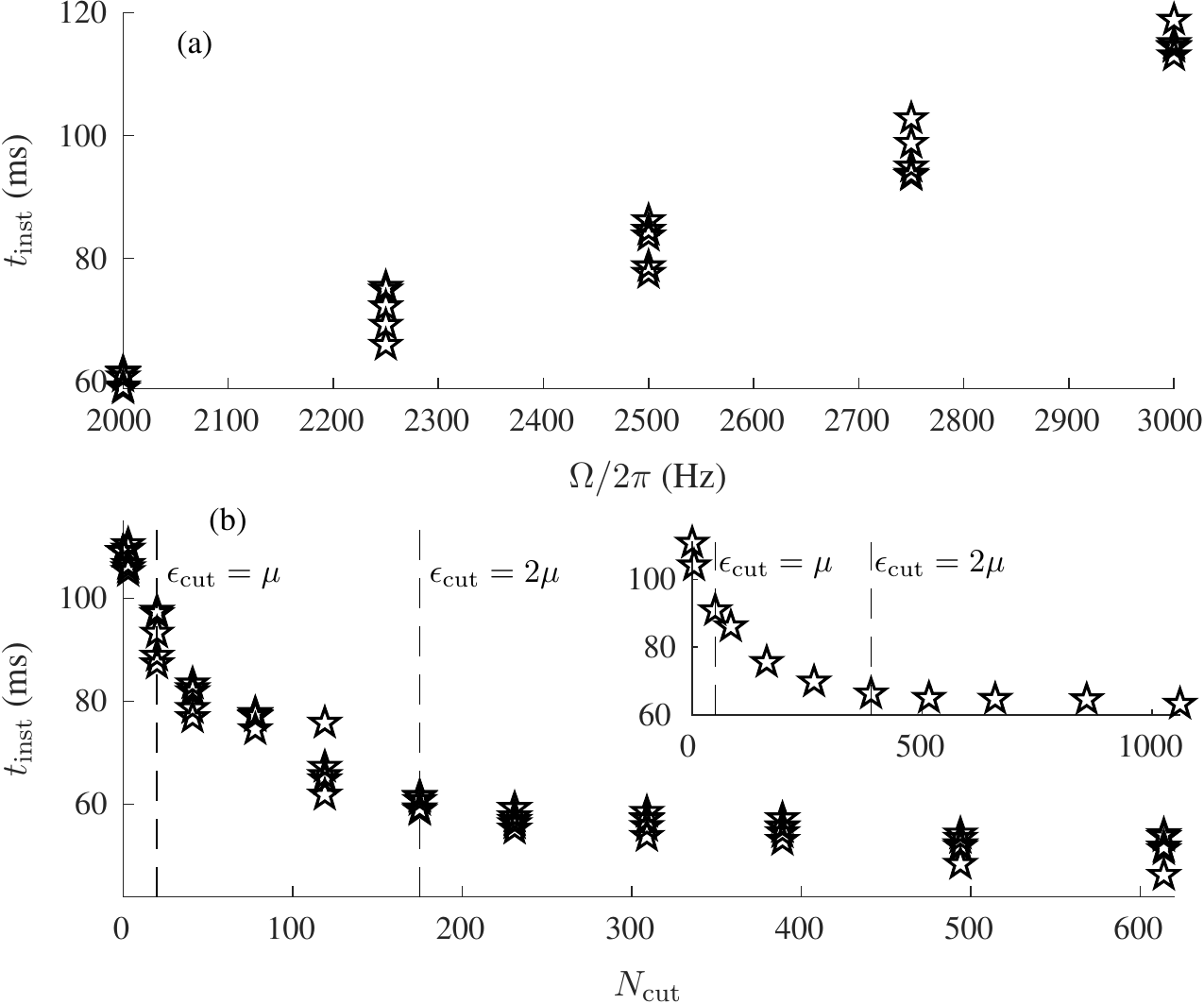}
   \caption{
  (a) Variation of instability time with trajectories. For each rotation frequency, we show $t_{\mathrm{inst}}$ for 5 trajectories seeded with independent initial state noise. Results here for $N_{\mathrm{cut}}=175$ with $\epsilon_{\mathrm{cut}}=2\mu$. (b) Dependence of the instability time on the amount of noise added. We show  $t_{\mathrm{inst}}$  as a function of the number of modes, $N_{\mathrm{cut}}$, that noise is added to. The two vertical dashed lines show where the associated single particle energy cutoff is equal to one and two times the chemical potential. The inset shows a similar calculation of the instability time for a condensate with $N_c=5\times10^3$ atoms.   Parameters are as in Fig.~\ref{fig:widthsED}(a).
   }
   \label{fig:instscaling}
\end{figure}

 \begin{figure}[htbp] 
   \centering
   \includegraphics[width=3.4in]{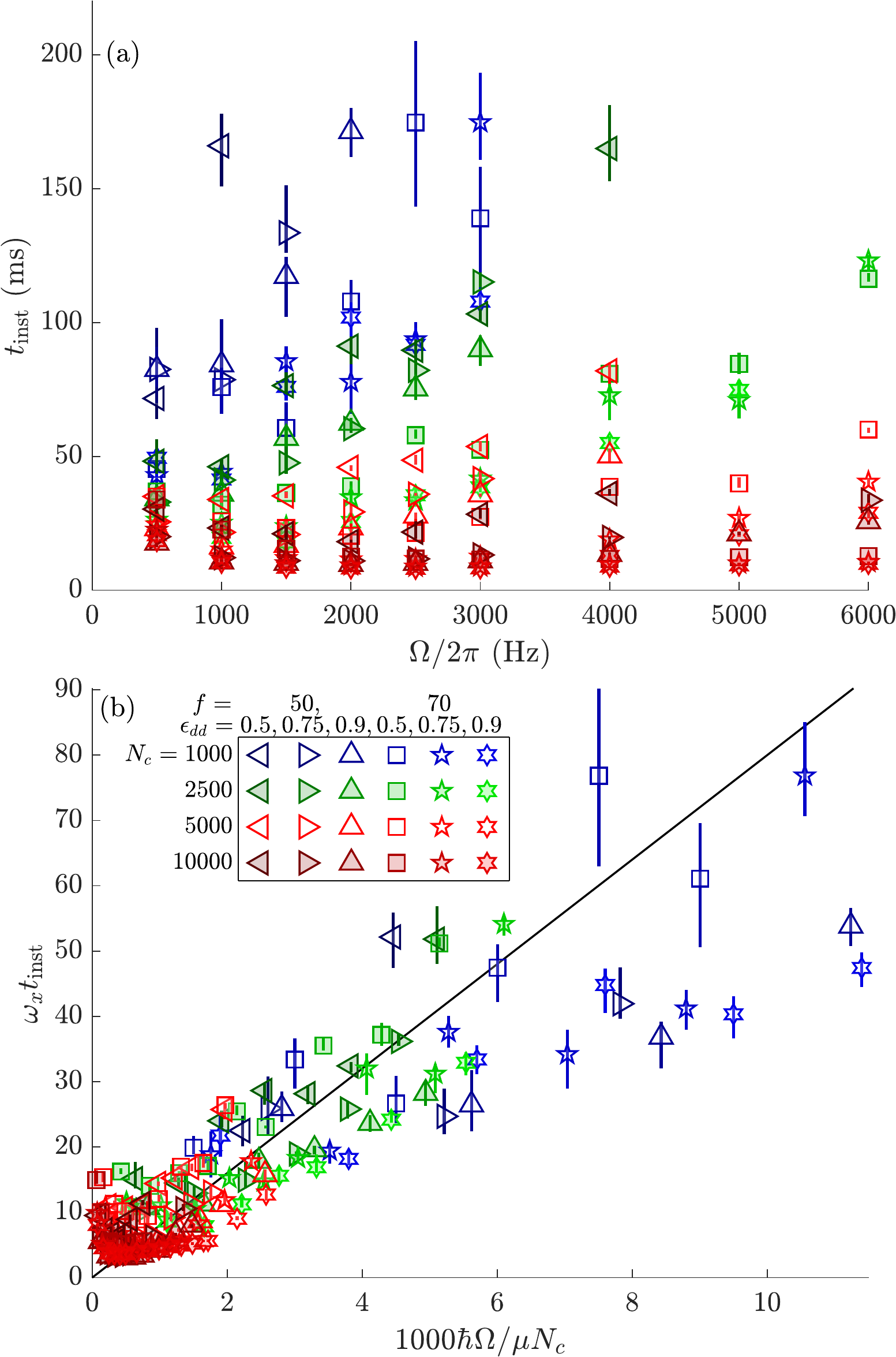}
   \caption{ (a) Bare and (b) scaled results for the instability time over a wide range of parameters.  The parameters for the simulations are indicated in legend of subplot (b), with the parameter $f$ specifying the trap frequencies as $(\omega_x,\omega_y,\omega_z)=2\pi f\times(1,1,1.1)$. The DDI tuning ramp is as indicated in Fig.~\ref{fig:widthsED}(a) and other parameters are the same as those used in Fig.~\ref{fig:widthsED}.
For each simulation case we take five trajectories to explore  stochastic effects of the noise. The means are shown by the symbols and the vertical lines show the range of results.  The slanted black line is a guide to the eye given by $\omega_x t_\mathrm{inst} = 8 \times 10^3\hbar\Omega/\mu N_c$. The chemical potential values, and other relevant quantities for the cases considered, are given in Appendix \ref{AppProps}.}
   \label{fig:collapse}
\end{figure}

To quantify the instability time, we define $t_{\mathrm{inst}}$ as the time when $E_K$ increases to twice its initial value. To tune the DDIs  successfully requires that $t_{\mathrm{inst}}$ is appreciably greater than $t_{\rm{on}}$. The dynamic instabilities are dependent on the sampling of noise in the initial state, and different trajectories will have different instability times. We show some examples of variance of $t_{\mathrm{inst}}$ for different trajectories in Fig.~\ref{fig:instscaling}(a). 

 We also examine the dependence of our results on the single particle energy cutoff $\epsilon_{\mathrm{cut}}$ used to define the modes in which the noise is added. Notably as  $\epsilon_{\mathrm{cut}}$   increases the number of modes with noise added grows as $N_{\mathrm{cut}}\propto \epsilon_{\mathrm{cut}}^2$, and hence the total noise added to the simulations increases. Our results in Fig.~\ref{fig:instscaling}(b) show that the instability time decreases as we add noise to higher energy modes (i.e.~as $N_{\mathrm{cut}}$ increases). However, once modes with a single particle energy of up to $\epsilon_{\mathrm{cut}}\approx2\mu$ are populated with noise, the instability time levels off and has a weak dependence on the noise.  A case with twice as many atoms ($N_c=5\times10^3$) and a higher chemical potential (see Appendix \ref{AppProps}) is shown in the inset to Fig.~\ref{fig:instscaling}(b), also confirming that the results converge for $\epsilon_{\mathrm{cut}}\approx2\mu$.
 This dependence on noise suggests that the most important modes for the instability occur in the energy range $\epsilon\in[0,2\mu]$, since populating higher energy modes with noise has negligible effect on the instabilities.

Finally, we explore the instability time over a wide parameter regime, varying the condensate atom number $N_c$, the ratio of the DDI to the contact interaction $\epsilon_{dd}$, the trap frequency $f$ (as defined in the caption of Fig.~\ref{fig:collapse}), and the rotation frequency $\Omega$.  For each system case $(N_c,\epsilon_{dd},f)$ we have simulated the tuning dynamics using five trajectories for each value of $\Omega$.  

We have checked that the results we show are independent of the choice of numerical simulation grid, as detailed in Appendix \ref{AppNum}. For sufficiently low rotation frequency (all of the results shown), the results are independent of the grid, i.e.~for different grids the changes in $t_{\mathrm{inst}}$ are smaller than 1\:ms. For higher rotation frequencies our simulations results become grid dependent and are not shown. In these high frequency simulations we necessarily must use small numerical time-steps (see Appendix \ref{AppNum}) and individual simulations in these cases take up to four days to run using Titan-V graphics processor unit (GPU) hardware. For faster rotation rates grid-converged results may be possible with larger grids, but the limited GPU memory means we are unable to thoroughly explore such cases at this time.  
 
 The results for the instability times of our simulations, shown in Fig.~\ref{fig:collapse}(a), are seen to vary strongly with parameters. The shortest instability times are $t_{\mathrm{inst}}\approx10\,$ms (i.e.~$< t_{\mathrm{on}}$) indicating the system becomes unstable before the conclusion of the tuning ramp. This case tends to occur for large atom numbers at low rotation frequencies. The instability time of the $N_c=5\times10^3$ and $f=70\,$Hz cases are seen to increase with $\Omega$, and we extrapolate that they would have a $100\,$ms instability time for $\Omega/2\pi\sim15\times10^3\,$Hz. The $N_c=10^4$ cases follow a similar trend, but increase more slowly with increasing $\Omega$.
 
 We find that our data can be scaled to better reveal the general trends. We utilize our earlier observation (related to the sensitivity of the instability time to the noise)  that the relevant instabilities appear to scale with an energy $\epsilon\sim\mu$, to identify $\mu$ as a key energy scaling parameter. Scaling the rotation frequency by $\mu N_c$, and the instability time by the trap frequency, the rescaled data is shown  Fig.~\ref{fig:collapse}(b).
 While the data does not completely collapse to a universal line, the general trend of instability time increasing with $\Omega$ is clearly revealed. Interestingly the $\mu N_c$ scaling of $\Omega$ suggests that the required rotation frequencies scale extensively with $N_c$, indicating that rotational tuning for large condensates will require large rotation frequencies, that may be challenging to obtain in experiments.

\section{Discussion and Conclusions}
In this paper we have simulated the process of rotational DDI tuning implemented using a realistic tuning ramp. We have examined the properties of the resulting state in dynamical simulations to verify the effects of the tuned interactions, and the ramp time scale needed to avoid excessively exciting collective modes. We observe the dynamic instabilities predicted by Prasad \textit{et al.}~\cite{Prasad2018a}, and find that the kinetic energy provides a useful observable of the instability and allows us to identify an instability time. Our results indicate that the time scale of the instabilities is sensitive to the rotation frequency of the magnetic field, and we find that in many cases the instability can be delayed for sufficiently high frequency.    
We emphasize that our results here are for the maximally tuned case, i.e.~with $\varphi=\pi/2$ where $\bar{g}_{dd}=-\frac{1}{2}g_{dd}$. In general we expect that the instabilities should be weaker for smaller values of $\varphi$.

The first experiments examining rotational tuning have observed a lifetime reduction, evidenced through the loss of atoms in the system. Our work suggests that a systematic study in terms of the kinetic energy evolution (or equivalently momentum width) will allow the instabilities to be systematically quantified. This may also help to understand to what extent the lifetime in experiment was due to magnetic field gradients versus the rotationally induced instability.

A deeper understanding of the microscopic origin of the instabilities is clearly needed. Prasad \textit{et al.}~\cite{Prasad2018a} presented the results of a polynomial basis approach that they used to quantify the eigenvalues of the linear excitations of the tuned condensate in the Thomas-Fermi approximation.  A non-zero imaginary part of an eigenvalue indicates that the associated mode is dynamically unstable, with the magnitude of the imaginary part relating to the growth rate. Our results motivate the need for a better understanding of the instabilities beyond the Thomas-Fermi approximation, and of the spatial character of the unstable modes. This may help identify factors that affect the instability timescale, and  suggest methods for extending the lifetime of condensates with rotationally tuned DDIs. 

Another important direction to consider is the effect of tuned DDIs in the regime of quantum droplets. Such droplets can occur when $\epsilon_{dd}>1$ (e.g., see \cite{Kadau2016a,Chomaz2016a,Schmitt2016a,Wachtler2016a,Wachtler2016b,Baillie2016b,Bisset2016a}) and the condensate becomes  unstable to mechanical collapse, but is stabilized by beyond meanfield terms, i.e.~leading order repulsive effect of quantum fluctuations. We have not included these quantum fluctuation effects in our results presented here, but they should have a minor effect  in the  $\epsilon_{dd}<1$ regime we have considered. In the case of a self-bound droplet, the chemical potential $\mu$ is typically negative (e.g.~see \cite{Baillie2017a}), and the scaling we predict in Fig.~\ref{fig:collapse}(b) clearly cannot hold.  

The results shown in this paper for dysprosium apply to other species with the same $\epsilon_{dd}$ and atom number, but with  $t_\mathrm{on}$, $t_\mathrm{inst}$, $\Omega$, and $f$ scaled by $ma_{dd}^2$. For $^{164}$Dy to $^{168}$Er, times must be divided by (and frequencies multiplied by) approximately $3.8$. The results of Fig.~\ref{fig:collapse}(b) are unchanged, but $f=50$ and $f=70\:\mathrm{Hz}$ for $^{164}$Dy correspond to $f\approx 190$ and $f\approx 267\:\mathrm{Hz}$ respectively for $^{168}$Er.

\begin{acknowledgments}
We thank S.~Prasad, B.~Mulkerin, A.~M.~Martin, Y.~Tang, L.~Chomaz, and B. Lev for helpful discussions.
We acknowledge the contribution of NZ eScience Infrastructure (NeSI) high-performance computing facilities, and support from the Marsden Fund of the Royal Society of New Zealand. \vspace{5mm}
\end{acknowledgments}

\appendix

\section{Numerical Treatment}\label{AppNum}
The initial state used in simulations is 
\begin{align}
    \psi(\bx,0) = \psi_0(\bx) + \sum_j{}^{'} \alpha_j \phi_j(\bx),
\end{align}
where $H_{sp}\phi_j = \epsilon_j\phi_j$, and $\alpha_j$ is a complex Gaussian noise with $\langle \alpha_j \rangle = 0$  and $\langle |\alpha_j|^2 \rangle = \tfrac12$. We restrict the sum to modes with energies 
$\epsilon_j \le \epsilon_{\mathrm{cut}}$, exclude the ground state, and denote the total number of modes within the sum as $N_{\mathrm{cut}}$.

We start with the dipoles aligned along $\zh$ and tilt into the $xy$-plane linearly while the dipoles rotate at frequency $\Omega$  using the ramp outlined in Section \ref{s:results} [see Fig.~\ref{fig:widthsED}(a)].
For calculating the dipolar interaction, we use a cutoff which is given by \eqref{e:Ur} for $r<R$ and zero otherwise. The resulting Fourier transform is \cite{Ronen2006a}
\begin{align}
    \UDt^R(\bk,t) = \UDt(\bk,t)\left[1 +3\frac{\cos(kR)}{(kR)^2}-3\frac{\sin(kR)}{(kR)^3}\right],
\end{align}
where
\begin{align}
    \UDt(\bk,t) =g_{dd}\left(3[\be(t)\cdot\kh]^2 -1\right).
\end{align}
For calculations shown in Section \ref{s:results}, we used numerical grids with $192$ gridpoints in each direction and a spatial grid resolution of $0.15\sqrt{\hbar/m\omega_x}$. For our grid checks, for each set of parameters and each rotation frequency, we also ran one of the trajectories on four additional grids, with (i) 50\% more gridpoints in the radial directions, (ii) 50\% more gridpoints in the axial direction, (iii) 50\% greater grid range in the radial directions, (iv) 50\% greater grid range in the axial direction. For one case, we also checked that if the results are independent of the simulation grid for one trajectory, then they are independent for all five trajectories. Our time step was $\Omega \Delta t = 0.01$. \vspace{5mm} 

\section{Simulation properties}\label{AppProps}
Here we provide additional details about the states and simulations used in Fig.~\ref{fig:collapse}. Table \ref{tab:mu} gives the trap parameter $f$ [recall $(\omega_x,\omega_y,\omega_z)=2\pi  f\times(1,1,1.1)$],  the relative dipole strength $\epsilon_{dd}$ (note we fix $a_{dd}=130.8\,a_0$),  and condensate number $N_c$. Also, the chemical potential $\mu$ for the initial condensate (without noise), the total number of atoms $N$ when half-an-atom of noise is added to single particle modes with energies $\epsilon_j\le2\mu$ and  $E/N$, where $E$ is the average total energy of the initial state including noise.\vspace{2mm}
\begin{table}[H]
   \centering
   \begin{tabular}{crr|rrr|r} %
       $f$ & $\epsilon_{dd}$ & $N_c$ & $N$ & $E/Nh$ & $\mu/h$ & $t_\mathrm{inst}$ \\
       (Hz) &  &  &  & (Hz) & (Hz) & (ms) \\\hline 
50 & 0.50 & 1000 & 1057 & 170 & 224 &  $>300$ \\
50 & 0.75 & 1000 & 1034 & 149 & 192 &  $>300$\\
50 & 0.90 & 1000 & 1028 & 140 & 178 &  $>300$ \\
70 & 0.50 & 1000 & 1060 & 251 & 333 & 139 \\
70 & 0.75 & 1000 & 1042 & 219 & 284 & 175 \\
70 & 0.90 & 1000 & 1028 & 205 & 263 & 108 \\ \hline
50 & 0.50 & 2500 & 2642 & 231 & 313 & 103 \\
50 & 0.75 & 2500 & 2588 & 198 & 264 & 115 \\
50 & 0.90 & 2500 & 2568 & 184 & 243 & 90 \\
70 & 0.50 & 2500 & 2677 & 344 & 467 & 52 \\
70 & 0.75 & 2500 & 2607 & 294 & 393 & 42 \\
70 & 0.90 & 2500 & 2582 & 272 & 361 & 38 \\ \hline
50 & 0.50 & 5000 & 5326 & 297 & 407 & 54 \\
50 & 0.75 & 5000 & 5194 & 252 & 342 & 42 \\
50 & 0.90 & 5000 & 5142 & 232 & 312 & 36 \\
70 & 0.50 & 5000 & 5394 & 443 & 608 & 27 \\
70 & 0.75 & 5000 & 5233 & 374 & 510 & 12 \\
70 & 0.90 & 5000 & 5177 & 344 & 466 & 12 \\ \hline
50 & 0.50 & 10000 & 10730 & 386 & 532 & 28 \\
50 & 0.75 & 10000 & 10429 & 324 & 445 & 13 \\
50 & 0.90 & 10000 & 10322 & 297 & 406 & 11 \\
70 & 0.50 & 10000 & 10884 & 576 & 796 & 11 \\
70 & 0.75 & 10000 & 10523 & 483 & 665 & 9 \\
70 & 0.90 & 10000 & 10390 & 442 & 606 & 8 \\ \hline
   \end{tabular}
   \caption{Details of Fig.~\ref{fig:collapse} cases. Here $N$ is the expected average total number of atoms in the field $\psi$ and  $E/N$ is the initial average energy per particle including noise up to the cutoff energy. The chemical potential $\mu$ is for the initial state $\psi_0$. The instability time is the average over five trajectories with $\Omega/2\pi = 3\:\mathrm{kHz}$. }
   \label{tab:mu}
\end{table}


%

\end{document}